\documentclass[aps,pra,reprint,nofootinbib]{revtex4-1}

\usepackage{amsmath}
\usepackage[amssymb]{SIunits}
\usepackage{url}
\usepackage{graphicx}
\usepackage{color}

\usepackage{soul}

\usepackage{hyperref}
\usepackage{mathtools}

\DeclarePairedDelimiter\abs{\lvert}{\rvert}%

\usepackage{lipsum}

\usepackage{xcolor}
\newcommand{\highlight}[1]{%
	\colorbox{black!10}{$\displaystyle#1$}}

\usepackage{footnote}
\usepackage{footmisc}

\DeclareMathOperator{\Tr}{Tr}

\begin{document}

\title{Theory of frequency modulated combs in lasers with spatial hole burning,\\ dispersion and Kerr}


\author{Nikola~\surname{Opa{\v c}ak}}
\author{Benedikt~\surname{Schwarz}}
\email{benedikt.schwarz@tuwien.ac.at}
\affiliation{Institute of Solid State Electronics, TU Wien, Gusshausstrasse 25-25a, 1040 Vienna, Austria}



\begin{abstract}
Frequency modulated (FM) frequency combs constitute an exciting alternative to generate equidistant spectra. The full set of Maxwell-Bloch equations is reduced to a single master equation for FM combs with fast dynamics to provide insight into the governing mechanisms behind phase-locking. It reveals that the recently observed linear frequency chirp is caused by the combined effects of spatial hole burning, group velocity dispersion and Kerr due to asymmetric gain. 
The comparison to observation in various semiconductor lasers suggests that the linear chirp is general to self-starting FM combs.
\end{abstract}

\keywords{quantum cascade detector, quantum well infrared photodetector, QWIP, mid-infrared, intersubband}
\maketitle

Optical frequency combs~\cite{haensch2006nobel,hall2006nobel} are lasers whose spectrum consists of a set of evenly spaced modes that obey a defined phase relation.
In the time domain, these lasers emit a signal, which, despite an eventual constant phase drift due to a non-zero carrier envelope offset frequency, is periodic.
In literature, frequency combs are mostly linked to ultra-fast lasers that emit short pulses. However, the Fourier theorem states that a comb spectrum is generated by any periodic signal, regardless of its shape.
A periodic frequency modulated (FM) signal is another example that fulfills this criterion. The first studies to generate such an FM laser output trace back to the 1960s, only a few years after the demonstration of the first laser~\cite{maiman1960stimulated}. An active intracavity phase modulator was used to generate FM oscillations in He-Ne \cite{Harris1964FM} and later in ND:YAG~\cite{Kuizenga1970FM} lasers.
Just from the similarity of the optical spectra to the Bessel amplitudes, it was concluded that FM lasers obey a sinusoidal modulation of the output frequency~\cite{Harris1965Theory,tiemeijer1989passive} and this picture remained dominant for over 50 years.

Today, FM combs experience a renaissance. In 2012, it was shown that quantum cascade lasers (QCLs) can be used to generate combs, whose intensity remains approximately constant~\cite{hugi2012mid}. In contrast to the work from the 1960s, the generated FM comb in QCLs is self-starting. 
The possibility of generating self-starting combs using the nonlinearity provided by the gain medium is particularly appealing for fundamental laser science  and the study of self-organization in complex nonlinear systems, but also of great interest for many applications.
FM combs can be generated in fast gain media, e.g. QCLs that do not satisfy the conditions for passive mode-locking~\cite{haus1975theory}. They are self-starting, requiring no additional components e.g. saturable absorbers, which is interesting for comb generation in interband cascade lasers (ICLs)~\cite{schwarz2018monolithic, bagheri2018passively}.
Both QCLs and ICLs emit in the mid-infrared region that is particularly appealing for dual-comb spectroscopy~\cite{villares2014dual,sterczewski2019mid}.

In this work, we provide a rigorous theoretical and numerical study of FM combs that is driven by recent experimental results \cite{singleton2018evidence,hillbrand2018coherent}. A highly optimized simulation tool was developed to reproduce the experimental results, to identify trends and to isolate the most relevant terms in the full set of nonlinear coupled differential equations. 
With this knowledge, we derive a simplified master equation for FM combs. It provides the eagerly awaited intuitive explanation of the phase-locking and answers the following questions:
\begin{itemize}
	\setlength\itemsep{0.em}
	\item[-] What triggers self-organization of the phases in FM combs to overcome chaos?
	\item[-] Why does the linear frequency chirp emerge from this competition, overcoming other solutions?
	\item[-] Why do QCL FM combs lock at low group velocity dispersion (GVD) \citep{Villares2016dispersion}?
	\item[-] Do FM combs require a fast gain medium?
\end{itemize}

 \begin{figure*}
 	\centering
 	\includegraphics[width = 1.\textwidth]{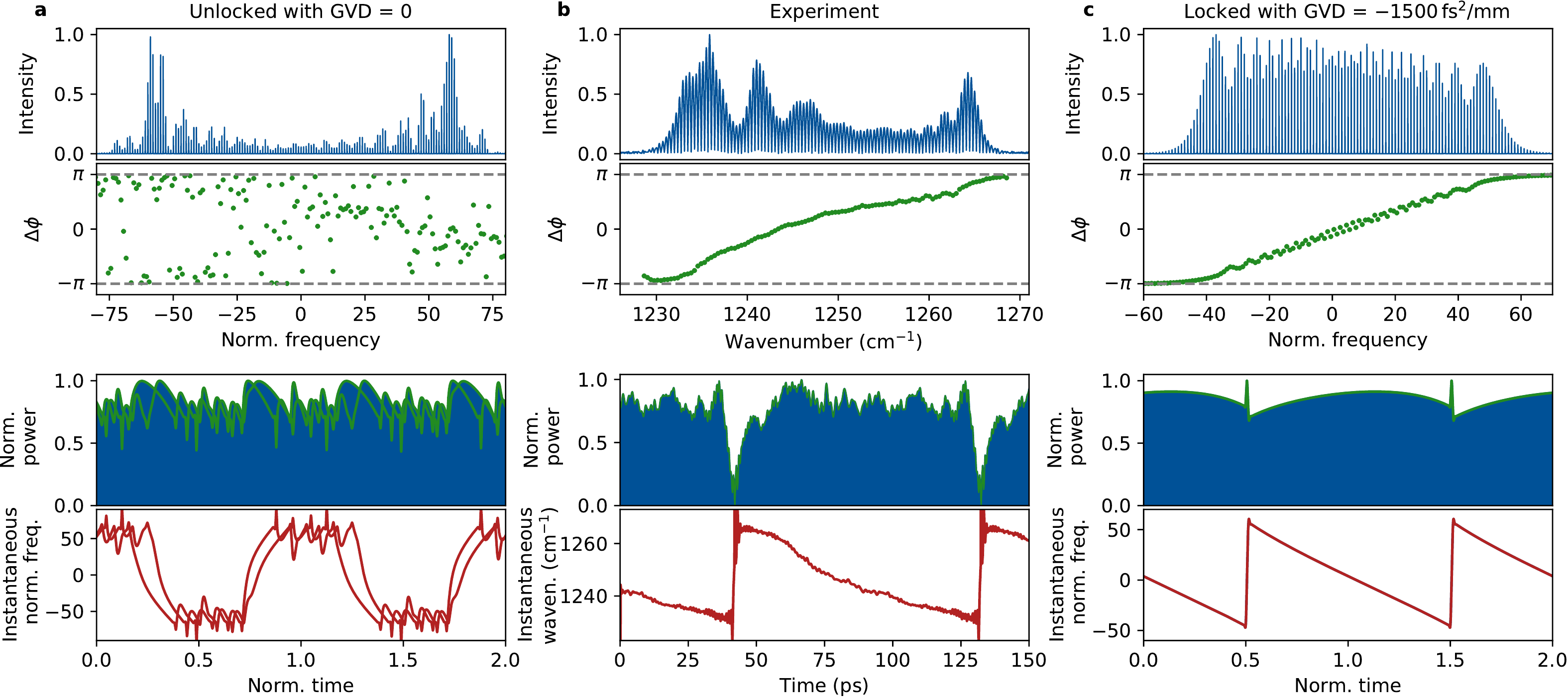}
 	\caption{ (a) Simulation of an unlocked laser. Top graph shows the normalized intensity spectrum with chaotic intermodal phases. Frequency is normalized to the round-trip frequency. Bottom graph shows the comparison of the time traces of the intensity and the normalized instantaneous frequency after 100 000 and 150 000 cavity round-trips. They do not match, indicating an unlocked state. (b) Experimental results, reprinted from~\citep{hillbrand2018coherent}, showing linear chirp. (c) Self phase-locked laser with non-zero GVD. Intermodal phases follow the linear pattern and compared time traces match perfectly. }
 	\label{fig1}
 \end{figure*}

The possibility to generate FM combs with QCLs is mostly explained through their fast gain dynamics and four-wave mixing via the occurrence of population oscillations that respond in anti-phase to oscillations of the light intensity\cite{Agrawal1988Population, Mansuripur2016Single}. A modulated intensity saturates the gain more than a constant intensity. Following the maximum emission principle~\cite{tang1967maximum}, amplitude modulations will be suppressed to maximize the output. 
In fast gain media, this effect is particularly strong. This is also the reason why a slow gain medium is required for pulse generation. There, the suppression is compensated and reversed by fast saturable absorption.

The main issue with this concept is that any phase arrangement that minimizes amplitude modulations is equal in energy, which should result in a chaotic phase modulation~\cite{Henry2017pseudorandom}. Figure \ref{fig1}a shows the corresponding numerical simulation result, reproducing the expected pseudo-random behavior. Experiments, however, clearly show the formation of a distinct periodic pattern and the generation of a frequency comb~\cite{singleton2018evidence,hillbrand2018coherent}. Figure \ref{fig1}b shows the experimental results of a QCL frequency comb with the characteristic linear phase pattern that covers the range from $-\pi$ to $\pi$. Note that we plot the intermodal phases, i.e. the phase difference between adjacent modes. This linear pattern corresponds to parabolic modal phases and a chirped instantaneous frequency.

The suppression of amplitude modulations can be interpreted as repulsive coupling of the intermodal phases.
The occurrence of self-organization in repulsively coupled systems can also be found in other fields of research, ranging from splay states in Josephson junctions~\cite{Strogatz1993splay} to cluster states in complex networks~\cite{Pecora2014cluster}. Such phenomena can be explained by additional contributions that induce an imbalance to favor one among many other solutions. 
Such an imbalance can for example be triggered by a finite GVD, as shown by our numerical results in figure~\ref{fig1}c.
This is particularly surprising, as the experimental observations of FM combs in QCLs were found in dispersion compensated cavities.
 
\begin{figure*}
	\centering
	\includegraphics[width = 1.\textwidth]{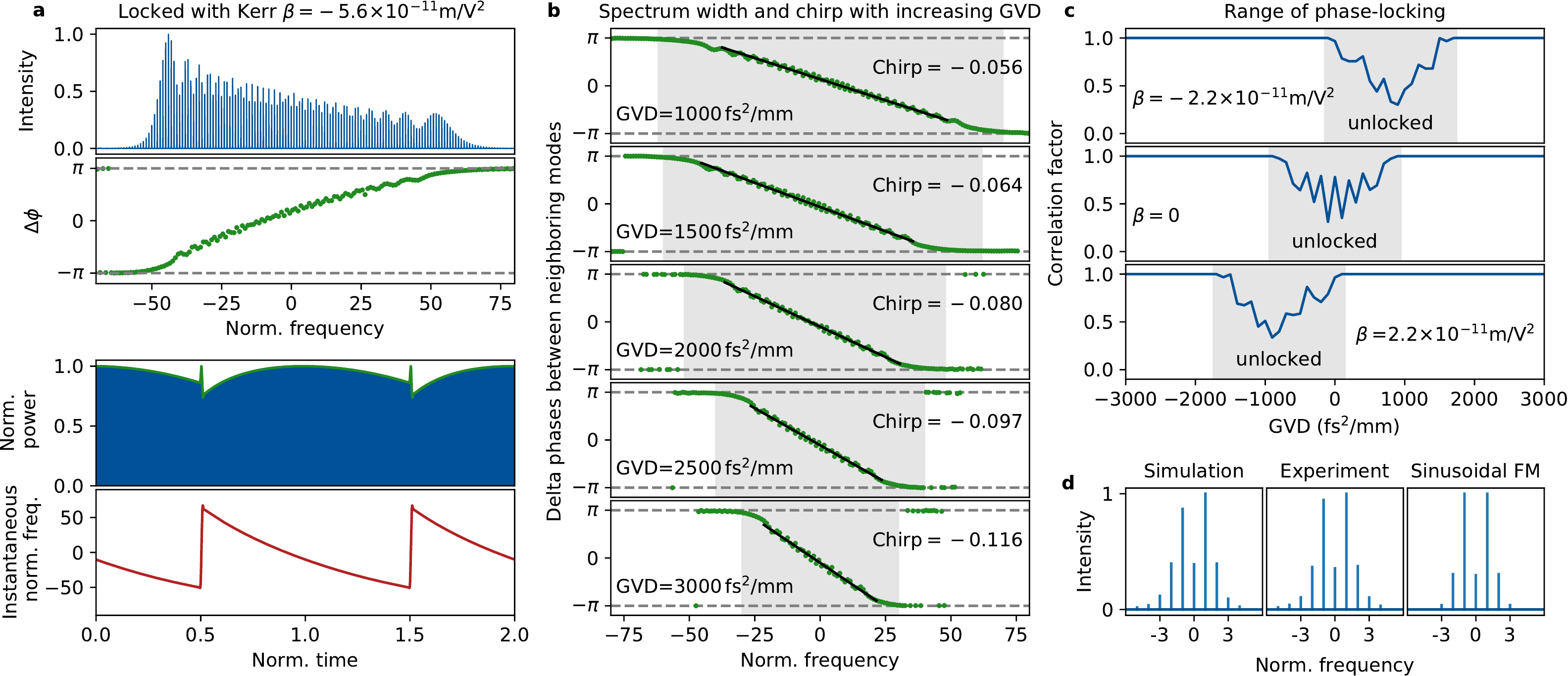}
	\caption{ (a) Simulation of a locked laser in the presence of Kerr. (b) Narrowing of the laser spectrum with an increasing GVD and the increase of the slope of the intermodal phases (chirp increase). Shaded regions indicate the spectrum width. (c) Shifting of the GVD value range required for self-locking, due to a change of Kerr. (d) Intensity spectrum comparison obtained from our simulation, experiment (recreated from\cite{tiemeijer1989passive}) and a sinusoidal FM given with the Bessel amplitudes. The comparison of the second harmonic spectra is shown in the supplementary material.}
	\label{fig2}
\end{figure*}

In the following, we explain why the GVD plays such a crucial role and which other effects are required to explain the experimental observations. The starting point is a system of eight coupled nonlinear differential equations that describes the manifold physics of a laser.
The system is based on the spatio-temporally resolved Maxwell-Bloch equations in the slowly varying envelope approximation~\cite{Shimoda1984Introduction, Wang2015Active}, which includes the effects of GVD and Kerr nonlinearity that have been mostly omitted previously. The details of the model and all derivations can be found in the supplementary material.
While this full model is capable of a quantitative analysis, it cannot provide an intuitive understanding of the underlying physics. 

Laser models with reduced complexity can yield an intuitive interpretation.
An example is the Haus master equation for mode-locking with saturable absorbers~\cite{haus1975theory}.
Such models utilize the adiabatic approximation to eliminate variables, e.g. the induced polarization and carrier populations.
However, with its application, the physics behind FM combs disappear.
In the adiabatic elimination the response of these variables is assumed to be instantaneous, which is equivalent to approximating their  transfer function by a constant, e.g. $H(\omega) = a/(1+i\omega T) \approx a$.
With this, the information on a remaining small phase delay that can accumulate over hundreds of round-trips is entirely lost.
In order to recover sufficient information, we use a Taylor expansion of the form $H(\omega) = a (1 - i\omega T- \omega^2 T^2...)$. 
This yields a single master equation for FM combs in fast gain media:
\begin{align}
\begin{split}
\Big( \frac{n}{c}&\partial_t  \pm \partial_z \Big) E_\pm = \frac{g(P)}{2} \Big[ E_\pm \highlight{- T_2 \partial_t E_\pm + T_2^2 \partial_t^2 E_\pm }\Big]\\
&-   \frac{g(P)T_g}{T_1 P_\mathrm{sat}} \Big[ \abs{E_\mp}^2 E_\pm \highlight{- ( T_2+T_\mathrm{g}) \abs{E_\mp}^2 \partial_t E_\pm} \\
&\highlight{-(T_2+T_\mathrm{g}) E_\pm E_\mp \partial_t E^*_\mp - T_2 E_\pm E^*_\mp \partial_t E_\mp } \Big]\\
&  + ik''\partial_t^2 E_{\pm} + i\beta \left( {\abs*{E_+}}^2\!+\!{\abs*{E_-}}^2 \right)E_{\pm} -\frac{\alpha_w}{2}  E_{\pm},
\label{eq:E_full} 
\end{split}
\end{align}
where $c$ is the speed of light, $n$ the refractive index, $E\pm$ the right and left propagating complex field envelopes, $T_1,T_2$ and $T_g$ the recovery times of the gain, polarization and the population grating, $k''$ the group velocity dispersion coefficient,  $\beta$ the Kerr nonlinearity coefficient, $\alpha_w$ the waveguide loss, $g(P) = g_0 / (1\!+\!P/P_\mathrm{sat})$ the saturated gain, $P_\mathrm{sat}$ the saturation power and $P = \abs{E_+}^2\!+\!\abs{E_-}^2$ the normalized power.
The highlighted terms correspond to the higher order terms introduced by the Taylor expansion. 
While we used this equation for our numerical simulations, we derive a reduced form to provide a better insight into the most relevant mechanisms.
  
Aiming for a qualitative explanation of FM combs, we rewrite eq.~\ref{eq:E_full} in terms of power and phase $E_\pm = \sqrt{P_\pm} \exp(i \phi_\pm)$ and neglect several minor contributions, e.g. terms with $\partial_t A$ are smaller than terms with $\partial_t\phi$.
The reduced master equation for FM combs reads:
\begin{align}
\begin{split}
\big( \frac{n}{c}\partial_t  &\pm \partial_z \big) P_\pm =  g(P) \Big[ P_\pm \highlight{ - \frac{T_2}{2} \partial_t P_\pm} \\
&\highlight{-  T_2^2 P_\pm (\partial_t \phi_\pm)^2 } \Big] - \alpha_w P_{\pm} , \label{eq:P_reduced}
\end{split}\\
\begin{split}
\Big( \frac{n}{c}\partial_t &\pm \partial_z \Big) \phi_\pm  =-k'' (\partial_t \phi_\pm)^2 +\beta \left( P_+\!+\!P_- \right) \\
&\:  
\highlight{+\frac{g(P)}{2} \Big[- T_2 \partial_t \phi_\pm
- \frac{T_g^2}{T_1 }\frac{P_\mp}{P_\mathrm{sat}} \partial_t \phi_\mp \Big] } .\label{eq:phi_reduced}
\end{split}
\end{align}
Now it will become clear why we introduced the Taylor expansion (highlighted terms) instead of the adiabatic elimination.
As a frequency modulation is essentially a modulation of the phase, FM comb operation is mostly governed by eq. \ref{eq:phi_reduced}. Without the highlighted terms the phase dynamics would be lost and with that also the physics of FM combs.
The first highlighted term in eq. \ref{eq:phi_reduced} dampens phase oscillations, tending towards single mode operation. The second highlighted term in eq. \ref{eq:phi_reduced} is due to spatial hole burning and
facilitates multi-mode operation. The dispersion term proportional to $k''$ determines the evolution of the cumulative phase shape and favors a convex or concave parabola, depending on the sign of $k''$. This results in a chirp in one direction or another.
The Magnitude of $k''$ is directly related to the chirp of the intermodal phases. If it is large enough such that the intermodal phases cover the full range of $2\pi$, the laser can produce a stable periodic output. A further increase makes the chirp larger which requires a narrowing of the spectrum (figure~\ref{fig2}b).
The residual amplitude modulation present in both the experiments and simulations (figure~\ref{fig1}) are mostly due to the second highlighted term in eq.~\ref{eq:P_reduced}.

The situation becomes slightly more complex, when considering an additional Kerr nonlinearity $\beta$. In that case, eq.~\ref{eq:P_reduced} and \ref{eq:phi_reduced} are dynamically coupled. The second highlighted term in eq. eq.~\ref{eq:P_reduced} represents a source term, forcing $P_\pm$ to oscillate in a similar manner. Coupling $P_\pm$ back in eq. \ref{eq:phi_reduced} through the Kerr term, one sees a similar effect as the GVD, as approximately both are influencing the phase through a term proportional to $(\partial_t \phi_\pm)^2$. Figure \ref{fig2}a shows the numerical simulation of an FM comb generated by a Kerr nonlinearity with zero GVD. One can observe a slight bending of the intermodal phases, also present in the experimental data (figure \ref{fig1}b).

The Kerr term mostly originates from a change of the real part of the refractive index with the population distribution. This is connected to an asymmetric spectral gain profile and commonly expressed through a non-zero linewidth enhancement factor (LEF)~\cite{Agrawal1988Population}.
A non-zero Kerr term or LEF strongly shift the range of GVD required for FM comb operation. This explains why QCL frequency combs have been found close to zero GVD.
Figure \ref{fig2}c shows the range of GVD required to obtain FM comb operation for three different values of $\beta$ that correspond to realistic values of LEF for QCLs at room-temperature \cite{Aellen2006direct}.
Altering the shape or width of the gain changes the required GVD and thus suggests considering this effect in the design of broadband FM combs.
It will be interesting to see if the observed behavior at very high values of GVD with the continuous narrowing of the spectrum can also be reproduced in the experiment, or if above a certain threshold the laser becomes unlocked again due to effects neglected in the reduced model. First attempts of solving the full model for high dispersion yielded again an unlocked state, but the results remain inconclusive due to numerical issues that are known from convection-diffusion problems. A detailed investigation will be part of future work.

We did not find a reason why FM comb operation should be strictly limited to fast gain media. 
In QCLs, with their fast dynamics, even a small gain asymmetry leads to a considerable Kerr nonlinearity.
In interband semiconductor lasers, this contribution is attenuated due to the slower dynamics, but the asymmetry is much more pronounced.
Moreover, dispersion driven FM comb operation appears to be independent of the gain dynamics.
Hence, we believe that FM comb operation with a linear chirp is a general phenomena.
As an example the presented theory also explains the experimental observations of self-starting FM oscillations in a InGaAsP laser diode\cite{tiemeijer1989passive}. The simulated intensity spectrum with the linear chirp fits the measured modal amplitudes much better than previously assumed Bessel amplitudes (figure~\ref{fig2}d).
Our theory can further explain observations in numerical simulations of quantum dot and quantum well lasers~\cite{gioannini2015time,dong2017traveling} and recent results on the demonstration of FM comb operation in interband cascade lasers\cite{schwarz2018monolithic}.

In conclusion, we provided detailed insights into the formation of frequency combs in single section lasers without saturable loss. Going beyond the adiabatic approximation, we derived a master equation to explain the physics and identified the most relevant contributions. In accordance to this, an FM comb requires: spatial hole burning to trigger multi-mode operation, gain saturation to suppress amplitude modulation and a minimum, but finite contribution from GVD or Kerr due to gain asymmetry that gives rise to a chirped output. 
A minimum amount is required such that the intermodal phases can cover a range of $2\pi$ over the spectral span to suppress amplitude modulations. Further increase will enforce a narrowing of the spectrum. 
The presented theory is capable of explaining experimental observations in various types of semiconductor lasers, indicating 
that the linear chirp is a general phenomenon behind the nature of FM combs.

This work was supported by the Austrian Science Fund (FWF) within the projects "NanoPlas" (P28914-N27) and "Building Solids for Function" (Project W1243).


%

\appendix

\section{General framework}
\subsection{Maxwell-Bloch equations }
We begin with the density matrix formalism for a two-level system \cite{Shimoda1984Introduction}. The total Hamiltonian of the system can be represented as $\hat{H}=\hat{H_0}+\hat{H'}$. Interaction of the system with the electric field is described with $\hat{H'}=-\hat{\mu} E(t)$, where $\hat{\mu}$ is the dipole moment operator. We can write:
\begin{align}
\hat{H}=
\left[ {\begin{array}{cc}
	W_l & -\mu_{lu}E \\
	-\mu_{ul}E & W_u \\
	\end{array} } \right]=
\left[ {\begin{array}{cc}
	\hbar \omega_l & -\mu_{lu}E \\
	-\mu_{ul}E & \hbar \omega_u \\
	\end{array} } \right],
\end{align}
where indices $l$ and $u$ denote lower and upper states and $W_{l,u}$ are the corresponding energies. The time evolution is governed with the von Neumann equation, with the time flow in direction $\sim e^{+i\omega t}$ :
\begin{align}
\begin{split}
&\frac{d\hat{\rho}}{dt}=\frac{i}{\hbar}[\hat H,\hat{\rho}]  = \frac{i}{\hbar}\\
&
\left[ {\begin{array}{lr}
	-\!\mu E(\rho_{ul}\!-\!\rho_{lu}) \hspace{1.2cm} -\!\rho_{lu}(\omega_u\!-\!\omega_l)\!-\!\mu E(\rho_{uu}\!-\!\rho_{ll}) \\
	\rho_{lu}(\omega_u\!-\!\omega_l)\!+\!\mu E(\rho_{uu}\!-\!\rho_{ll})  \hspace{1.8cm} \mu E(\rho_{ul}\!-\!\rho_{lu}) \\
	\end{array} } \right],
\end{split}  
\end{align}
bearing in mind that $\rho_{ul}={\rho^{\ast}_{lu}}$. Let us also denote the transition frequency $\omega_0=\omega_u-\omega_l$. Since the matrix elements $\rho_{ll,uu}$ represent occupation probabilities for the levels, by multiplying them with the total sheet density $n_{tot}$, we obtain equations for the surface densities of the levels $n_l$ and $n_u$. We also normalize in the same way $n_{ul}=n_{tot}\rho_{ul}$. Furthermore, we add the dephasing processes and transitions, modeled with the polarization dephasing time $T_2$ and different transition lifetimes. Carrier diffusion is introduced via the coefficient $D$ and a pumping current $J$ to the upper level is also included. The pumping current is normalized to the elementary charge. The new set of equations are:
\begin{flalign}
\begin{split}
&\frac{\partial n_l}{\partial t}=\frac{n_u}{T_{ul}}-\frac{n_l}{T_{lg}}
+2\frac{\mu E}{\hbar}\operatorname{Im}(n_{ul})+D\frac{\partial^2 n_l}{\partial t^2},\\
&\frac{\partial n_u}{\partial t}=J-n_u(\frac{1}{T_{ul}}-\frac{1}{T_{ug}})
-2\frac{\mu E}{\hbar}\operatorname{Im}(n_{ul})+D\frac{\partial^2 n_u}{\partial t^2}, \\
&\frac{\partial n_{ul}}{\partial t}=(i\omega_0-\frac{1}{T_2})n_{ul}
+i\frac{\mu E}{\hbar}(n_u-n_l),
\label{eq:n}
\end{split}
\end{flalign}
where the index $g$ stands for the ground level and $T_{ij}$ represents the transition lifetime between levels $i$ and $j$. Pumping current $J$ is assumed to be constant over the entire laser cavity. In the next step, we find the macroscopic polarization $P$, where $L$ is the thickness of the doped region:
\begin{align}
P=\frac{n_{tot}}{L}\Tr [\hat{\rho} \hat{\mu}]=
\frac{\mu}{L}(n_{ul}+n^{\ast}_{ul}),
\label{eq:polar}
\end{align}
Employing the last relation, we can write the wave equation, where $\Gamma$ is the confinement factor and $n$ is the refractive index:
\begin{align}
\frac{\partial^2 E}{\partial z^2}-\frac{n^2}{c^2}\frac{\partial^2 E}{\partial t^2}=
\frac{\Gamma \mu}{\varepsilon_0 c^2L}\frac{\partial^2 }{\partial t^2}(n_{ul}+n^{\ast}_{ul}),
\label{eq:max}
\end{align}

We can now express the electric field $E(z,t)$ as a sum of backward and forward propagating components, which allows us to do the same with $n_{ul}$. Furthermore, all populations contain a grating component besides a spatial independent average component in order to account for the spatial hole burning (SHB). This ansatz is summed up in:
\begin{flalign}
\begin{split}
&E(z,t)=\!\frac{1}{2}[E_+(z,t)e^{i(\omega_0 t-k_0z)}\!+\!E_-(z,t)e^{i(\omega_0 t+k_0z)}\!+\!c.c.], \\
&n_{ul}(z,t)=\sigma_+e^{i(\omega_0 t-k_0z)}+\sigma_-e^{i(\omega_0 t+k_0z)}, \\
&n_l(z,t)=n_{l0}+n_{l2}e^{-2ik_0z}+n^{\ast}_{l2}e^{2ik_0z}, \\
&n_u(z,t)=n_{u0}+n_{u2}e^{-2ik_0z}+n^{\ast}_{u2}e^{2ik_0z}.
\label{eq:ansatz}
\end{split}
\end{flalign}

Substitution of expressions (\ref{eq:ansatz}) in the density matrix equations (\ref{eq:n}) and the wave equation (\ref{eq:max}) yields the equations for the envelope functions, after applying the slowly varying envelope and rotating wave approximations and adding the intensity loss coefficient of the waveguide $\alpha_w$ to the field envelope equations. Additionally, the Kerr effect is included and modeled as a phase variation through the Kerr coefficient $\beta$, so we can write:
\begin{flalign}
&\frac{\partial n_{l0}}{\partial t}=\frac{n_{u0}}{T_{ul}}-\frac{n_{l0}}{T_{lg}}
-\frac{\mu}{\hbar}\operatorname{Im}(E_+\sigma^{\ast}_++E_-\sigma^{\ast}_-),\label{eq:nl0} \\
&\frac{\partial n_{l2}}{\partial t}=\frac{n_{u2}}{T_{ul}}-\frac{n_{l2}}{T_{lg}}
+i\frac{\mu}{2\hbar}(E_+\sigma^{\ast}_--E^{\ast}_-\sigma_+)-4k^2Dn_{l2},
\label{eq:nl2} \\
&\frac{\partial n_{u0}}{\partial t}=J-(\frac{1}{T_{ul}}+\frac{1}{T_{ug}})n_{u0}
+\frac{\mu}{\hbar}\operatorname{Im}(E_+\sigma^{\ast}_++E_-\sigma^{\ast}_-),\label{eq:nu0} \\
&\frac{\partial n_{u2}}{\partial t}=-(\frac{1}{T_{ul}}+\frac{1}{T_{ug}})n_{u2}
-i\frac{\mu}{2\hbar}(E_+\sigma^{\ast}_--E^{\ast}_-\sigma_+) \notag \\
&\qquad\quad\enspace -4k^2Dn_{u2},
\label{eq:nu2} \\
&\frac{\partial \sigma_{+}}{\partial t}=-\frac{\sigma_+}{T_{2}}
+i\frac{\mu}{2\hbar}[E_+(n_{u0}-n_{l0})+E_-(n_{u2}-n_{l2})], 
\label{eq:sigma+} \\
&\frac{\partial \sigma_{-}}{\partial t}=-\frac{\sigma_-}{T_{2}}
+i\frac{\mu}{2\hbar}[E_-(n_{u0}-n_{l0})+E_+(n^{\ast}_{u2}-n^{\ast}_{l2})],
\label{eq:sigma-} \\
&(\frac{n}{c}\frac{\partial}{\partial t}+\frac{\partial}{\partial z})E_+=-i\frac{\Gamma \mu \omega_0}{n\varepsilon_0 c L}\sigma_+ +i\beta( {\abs*{E_+}}^2+{\abs*{E_-}}^2 )E_+ \notag \\
&\qquad\qquad\qquad\qquad  -\frac{\alpha_w}{2}E_+,
\label{eq:E+} \\
&(\frac{n}{c}\frac{\partial}{\partial t}-\frac{\partial}{\partial z})E_-=-i\frac{\Gamma \mu \omega_0}{n\varepsilon_0 c L}\sigma_- +i\beta( {\abs*{E_+}}^2+{\abs*{E_-}}^2 )E_- \notag \\
&\qquad\qquad\qquad\qquad  -\frac{\alpha_w}{2}E_-.
\label{eq:E-} 
\end{flalign}

Equations (\ref{eq:nl0})-(\ref{eq:E-}) make a complete set of coupled spatio-temporal density matrix and Maxwell equations similar to the ones used in \cite{Wang2015Active}.

\subsection{Group velocity dispersion}
In reality, a laser cavity always possesses a non-zero dispersion which has a profound impact on the intermode dynamics of the laser. Hence, if one aims for modeling the delicate process of phase-locking of the laser, it is of interest to obtain a time-domain wave equation that could fully describe the evolution of the electromagnetic field inside the cavity. We start from a one-dimensional wave equation that considers the possibility of dispersion:

\begin{align}
\frac{\partial^2 E(z,t)}{\partial z^2}\!-\!\frac{1}{c^2}\frac{\partial^2}{\partial t^2} \int_{-\infty}^{t}\!\!\!\!\!\!\varepsilon(t-\tau)E(z,\tau)d\tau\!=\!
\frac{1}{\varepsilon_0 c^2 }\frac{\partial^2 P(z,t)}{\partial t^2},\label{eq:7} 
\end{align}
where $P$ stands for the nonlinear macroscopic polarization and the dielectric permittivity $\varepsilon(t)$ is calculated as:
\begin{align}
\varepsilon(t)=\frac{1}{2\pi}\int_{-\infty}^{t}\varepsilon(\omega)e^{i\omega t}d\omega.
\end{align}
The convolution integral in equation (\ref{eq:7}) is not convenient for calculation, so we will try to obtain a more appropriate form. After applying the Fourier transform to equation (\ref{eq:7}), knowing the identity $\mathcal{F}(\frac{\partial^n}{\partial t^n}x(t))=(i\omega)^nX(\omega)$, we obtain the exact form of the equation in the frequency domain, where $\omega$ stands for the instantaneous frequency:
\begin{align}
\frac{\partial^2 E(z,\omega)}{\partial z^2}+\frac{\omega^{2}}{c^2}\varepsilon(\omega)E(z,\omega)=
-\frac{\omega^2}{\varepsilon_0 c^2 }P(z,\omega).\label{eq:8} 
\end{align}
We can now introduce the wave number $k(\omega)=\omega n(\omega)/c$, where $\varepsilon(\omega)=n^2(\omega)$. Also, let $\omega_0$ be the carrier frequency, which is introduced in the ansatz (\ref{eq:ansatz}) and $k_0 = k(\omega_0)$. From (\ref{eq:8}) we then obtain:
\begin{align}
\frac{\partial^2 E(z,\omega)}{\partial z^2}+k^2(\omega)E(z,\omega)=
-\frac{\omega^2}{\varepsilon_0 c^2 }P(z,\omega).\label{eq:9} 
\end{align}
Knowing that the instantaneous frequency $\omega$ is in the vicinity of $\omega_0$, we can write the Taylor expansion of $k(\omega)$:
\begin{align}
k(\omega)=k(\omega_0)+\frac{dk}{d\omega}\biggr|_{\omega_0}(\omega-\omega_0)+\sum_{m=2}^{+\infty} \frac{k_0^{(m)}}{m!}(\omega-\omega_0)^m,\label{eq:10} 
\end{align}
where $k_0^{(m)}=\partial^m k/\partial \omega^m$ and the group velocity can be defined as $v_g=1/(dk/d\omega)|_{\omega_0}\approx c/n$. Now it is a good idea to insert ansatzes (\ref{eq:ansatz}) in equation (\ref{eq:8}) and after some calculation, equations for the envelope functions can be derived:
\begin{align*}
\frac{1}{2} \bigg(\frac{\partial^2 }{\partial z^2} \mp 2ik_0\frac{\partial}{\partial z}-k_0^2 \bigg)E_{\pm} +\frac{1}{2} k^2(\omega)E_{\pm}=
-\frac{\omega^2\Gamma \mu}{\varepsilon_0 c^2 L }\sigma_{\pm},
\end{align*}
which can also be written as:
\begin{align}
&\frac{1}{2}\frac{\partial^2 E_{\pm} }{\partial z^2}    \mp ik_0\frac{\partial E_{\pm} }{\partial z} =-k_0(k-k_0)E_{\pm}   -\frac{1}{2}(k-k_0)^2E_{\pm} \notag \\
&\qquad\qquad\qquad\qquad\quad\:\,  -\frac{\omega^2\Gamma \mu}{\varepsilon_0 c^2 L }\sigma_{\pm}.     \label{eq:11} 
\end{align}
Keeping in mind that $E, n_ul \sim e^{i\omega t}$, we can conclude that $E_{\pm}, \sigma_{\pm} \sim e^{i(\omega-\omega_0) t}$. This means that $(\omega-\omega_0)\{E_{\pm},\sigma_{\pm}\}=-i\partial/\partial t\{E_{\pm},\sigma_{\pm}\}$. After inserting this in equation (\ref{eq:10}), we get:
\begin{align}
k=k_0+\frac{n}{c}(-i\frac{\partial}{\partial t})+\sum_{m=2}^{+\infty} \frac{k_0^{(m)}}{m!}(-i\frac{\partial}{\partial t})^m.\label{eq:12} 
\end{align}
Combining expressions  (\ref{eq:11}) and (\ref{eq:12}) one can obtain:
\begin{align*}
\frac{1}{2}\frac{\partial^2 E_{\pm} }{\partial z^2}    \!\mp\! ik_0\frac{\partial E_{\pm} }{\partial z} \!=\! -k_0 \bigg(\!\!-\!\frac{n}{c}i\frac{\partial }{\partial t} \!+\!\! \sum_{m=2}^{+\infty} \frac{k_0^{(m)}}{m!}(-i\frac{\partial}{\partial t})^m                                                                                                                                                                                                                                                                                                                                                                                                                                                                                                              \bigg)  E_{\pm}       \\
-\frac{1}{2}\bigg(-\frac{n}{c}i\frac{\partial }{\partial t} + \sum_{m=2}^{+\infty} \frac{k_0^{(m)}}{m!}(-i\frac{\partial}{\partial t})^m                                                                                                                                                                                                                                                                                                                                                                                                                                                                                                              \bigg)^2E_{\pm} \\
-\frac{\Gamma \mu}{\varepsilon_0 c^2 L }(\omega_0-i\frac{\partial }{\partial t})^2\sigma_{\pm}. 
\end{align*}
After some derivation and disregarding all derivatives of the same or higher order than $\mathcal{O}\Big( \frac{\partial^3 }{\partial t^3} \Big)$ in the above equation, one gets:
\begin{align*}
&\frac{i}{2k_0}\Big(\frac{\partial^2 }{\partial z^2}-\frac{n^2}{c^2}\frac{\partial^2 }{\partial t^2}\Big)E_{\pm}+	\Big(\pm\frac{\partial }{\partial z}+\frac{n}{c}\frac{\partial }{\partial t}\Big)E_{\pm}\\
&-\underbrace{i\frac{k_0^{(2)}}{2}\frac{\partial^2E_{\pm} }{\partial t^2}}_\text{$dispersion$}
=-i\frac{\Gamma \mu}{\varepsilon_0 c^2 L k_0}\Big(\omega_0^2-2i\omega_0\frac{\partial }{\partial t}-\frac{\partial^2 }{\partial t^2}\Big)\sigma_{\pm}.
\end{align*}

Looking at the last expression, we can neglect the second order derivatives in terms of both time and space in the first bracket for two reasons. Namely, they come with opposite signs and practically cancel out in addition to being minor in value, $\frac{\partial^2 }{\partial z^2}<<k_0\frac{\partial }{\partial z}$ and $\frac{\partial^2 }{\partial t^2}<<\omega_0\frac{\partial }{\partial t}$. Furthermore, both the first and second order time derivative in the bracket on the right-hand side of the equation can be neglected since the induced polarization $\sigma$ is a perturbation and is small compared to $\varepsilon_0 E$. Lastly, we can single out the term $i\frac{k_0^{(2)}}{2}\frac{\partial^2E_{\pm} }{\partial t^2}$ as the term which describes the group velocity dispersion (GVD) in the cavity. This is a straightforward step, keeping in mind the fact that we have obtained this term from the Taylor expansion of the wavevector around the central frequency. One should note that the actual value of the GVD is given with $k_0^{(2)}$. For simplicity, we will introduce the dispersion coefficient as $k''=\frac{k_0^{(2)}}{2}$ and also include  the waveguide losses and the Kerr effect as we did previously in equations (\ref{eq:E+}) and (\ref{eq:E-}). The last equation then transforms to the final two equations for the evolution of the forward and backward propagating components of the electric field:
\begin{flalign}
&(\frac{n}{c}\frac{\partial}{\partial t} \pm \frac{\partial}{\partial z})E_\pm -ik''\frac{\partial^2E_\pm }{\partial t^2}=-i\frac{\Gamma \mu \omega_0}{n\varepsilon_0 c L }\sigma_\pm \notag \\
&\qquad\qquad\qquad  +i\beta( {\abs*{E_+}}^2+{\abs*{E_-}}^2 )E_\pm -\frac{\alpha_w}{2}E_\pm,
\label{eq:13}
\end{flalign}
Equations (\ref{eq:nl0}) - (\ref{eq:sigma-}) combined with the wave equations (\ref{eq:13}) make a complete system that is used in this work.

\section{Frequency modulated comb - master\\equation}
Based on the recent experimental data that can be found in the main body of the paper, self phase-locked lasers give rise to frequency modulated (FM) frequency combs with a particular linearly chirped frequency output. The aim of this work is to try to shed some light on the mechanisms that are responsible.
As already mentioned, equations (\ref{eq:nl0}) - (\ref{eq:sigma-}) and (\ref{eq:13})) give an accurate quantitative model that is valid in the general case. However, they represent a coupled system of differential equations. Such a system is not very helpful if one aims for acquiring an intuitive understanding of the laser intermode dynamics that lead to a formation of an FM frequency comb. Furthermore, changing one parameter influences several equations and it is not clear what is the dominant effect that emerges as a consequence. Hence, it is preferable if it is possible to eliminate some of the equations and simplify the system. We will do so by considering that the gain medium possesses fast dynamics. This will allow us to obtain a model based only on the wave equation  containing additional terms - a master equation. For a fast gain medium, it serves as a quantitative tool almost as good as the full model provided previously. However, concerning the qualitative abilities, it is far more superior, giving insight into how different terms affect the process of self phase-locking of the laser. 

We can start with neglecting the lower level population, assuming that the extraction from the lower level is efficient, as well as the pumping to the upper level. Lifetimes $T_{1}$ and $T_{g}$ can be introduced for the static population $n_{u0}$ and the grating part of the population $n_{u2}$, respectively, as:
\begin{flalign}
&T_{1}= {\Big( \frac{1}{T_{ul}} + \frac{1}{T_{ug}} \Big)}^{-1},
\label{eq:Tnu0} \\
&T_{g}= {\Big( \frac{1}{T_{ul}} + \frac{1}{T_{ug}} +4k^2D \Big)}^{-1}.
\label{eq:Tnu2} 
\end{flalign}
Upon dividing the polarization into contributions from $n_{u0}$ and $n_{u2}$ as $\sigma_{\pm}=\sigma_{0\pm}+\sigma_{2\pm}$,the system can be written as:
\begin{flalign}
&\frac{\partial n_{u0}}{\partial t}=J- \frac{n_{u0}}{T_{1}}
+\frac{\mu}{\hbar}\operatorname{Im}(E_+\sigma^{\ast}_++E_-\sigma^{\ast}_-),
\label{eq:nu0v1} \\
&\frac{\partial n_{u2}}{\partial t}=- \frac{n_{u2}}{T_{g}}
-i\frac{\mu}{2\hbar}(E_+\sigma^{\ast}_--E^{\ast}_-\sigma_+),
\label{eq:nu2v1} \\
&\frac{\partial \sigma_{0\pm}}{\partial t}=-\frac{\sigma_{0\pm}}{T_{2}}
+i\frac{\mu}{2\hbar}E_{\pm}n_{u0}, 
\label{eq:sigma0+-} \\
&\frac{\partial \sigma_{2\pm}}{\partial t}=-\frac{\sigma_{2\pm}}{T_{2}}
+i\frac{\mu}{2\hbar}E_{\mp}n^{(*)}_{u2}, 
\label{eq:sigma2-v1} \\
&\frac{n}{c}\frac{\partial E_{\pm}}{\partial t} \pm \frac{\partial E_{\pm}}{\partial z} -ik''\frac{\partial^2E_{\pm} }{\partial t^2} =-i\frac{\Gamma \mu \omega_0}{n\varepsilon_0 c L}\sigma_{\pm} \notag \\
&\qquad\qquad\qquad +i\beta( {\abs*{E_+}}^2+{\abs*{E_-}}^2 )E_{\pm} -\frac{\alpha_w}{2}E_{\pm}.
\label{eq:E+-v1} 
\end{flalign}

Let us first eliminate $n_{u0}$. Its value is strictly real and has no contribution to the the phases of the field envelopes, so it does not affect the process of phase locking directly. Hence, one could expect that it is enough to calculate it from equation (\ref{eq:nu0v1}) by simply employing the adiabatic approximation, e.g. setting the time derivative to zero. We could conclude since in the FM comb, dynamics of the modal phases is much more significant than the dynamics of the modal amplitudes. We can eliminate the polarization $\sigma_{\pm}$ by just considering the $\sigma_{0\pm}$ contribution which we calculate after making the adiabatic approximation in equation (\ref{eq:sigma0+-}). Hence, after replacing $\sigma_{\pm} = i\frac{\mu T_2}{2 \hbar}n_{u0}E_{\pm}$ in the equation (\ref{eq:nu0v1}), we have, after some calculation:
\begin{align}
n_{u0} = \frac{ T_{1} J}  {  1 + \frac{{\abs*{E_+}}^2+{\abs*{E_-}}^2} {E^2_{sat}}  }, 
\label{eq:nu0v2} 
\end{align}
where the saturation field $E_{sat}$ has been introduced as $E^2_{sat}=2\hbar^2/(\mu^2T_1 T_{2})$.

Let us now analyze equation (\ref{eq:nu2v1}). After applying the Fourier transform, bearing in mind that $\mathcal{F}(\frac{\partial}{\partial t})=i(\omega-\omega_0)$ since we are dealing with the envelope functions, we can write:
\begin{align*}
n_{u2}& = \frac{1}{ 1 +i(\omega-\omega_0)T_{g}} ( -i\frac{\mu T_{g}}{2\hbar} ) \mathcal{F} \Big(E_+\sigma^{\ast}_--E^{\ast}_-\sigma_+ \Big)  \\
&\approx  -i\frac{\mu T_{g}}{2\hbar} (1 -i(\omega-\omega_0)T_{g}) \mathcal{F} \Big(E_+\sigma^{\ast}_--E^{\ast}_-\sigma_+ \Big), 
\end{align*}
so that after applying the inverse Fourier transform:
\begin{align}
n_{u2} = -i\frac{\mu T_{g}}{2\hbar} (1 -T_{g}\frac{\partial}{\partial t})  (E_+\sigma^{\ast}_--E^{\ast}_-\sigma_+ ). 
\label{eq:15} 
\end{align}
We have utilized the assumption that the gain medium is fast in the derivation of the last relation when we made the approximate step in the Taylor expansion of $( 1 +i(\omega-\omega_0)T_{n_{u2}})^{-1}$. We can again eliminate the polarization $\sigma_{\pm}$ by replacing $\sigma_{\pm} = i\frac{\mu T_2}{2 \hbar}n_{u0}E_{\pm}$ in the equation (\ref{eq:15}):
\begin{align}
n_{u2} = -\frac{\mu^2 T_g T_{2}}{2{\hbar}^2} n_{u0}  \Big(E_+E^{\ast}_- - T_{g}\frac{\partial E_+}{\partial t}E^{\ast}_-  -   T_{g} E_+\frac{\partial E^{\ast}_- }{\partial t} \Big ). 
\label{eq:nu2v2} 
\end{align}
In the last equation it was possible to take $n_{u0}$ outside of the brackets since the value is strictly real and hence its derivative would not affect the phases of the field envelopes.

A similar procedure can be done concerning $ \sigma_{0\pm}$. After applying the Fourier and inverse Fourier transform and also keeping terms up to $\mathcal{O}\Big( \frac{\partial^2 }{\partial t^2} \Big)$ in the Taylor expansion, one can obtain:
\begin{align*}
\sigma_{0\pm} = i\frac{\mu T_{2}}{2\hbar} n_{u0}  \Big(E_{\pm} - T_{2}\frac{\partial E_{\pm}}{\partial t} + T^2_{2}\frac{\partial^2 E_{\pm}}{\partial t^2}\Big). 	
\end{align*}
The reason behind keeping the second derivative in the Taylor expansion will be clear afterwards. After combining equation (\ref{eq:nu0v2}) with the previous relation, the expression for $ \sigma_{0\pm}$ is obtained:
\begin{align}
\sigma_{0\pm} = i\frac{\mu T_{1}T_{2} J }{2\hbar \Big( 1 + \frac{{\abs*{E_+}}^2+{\abs*{E_-}}^2} {E^2_{sat}}  \Big) }  \Big(E_{\pm} - T_{2}\frac{\partial E_{\pm}}{\partial t} + T^2_{2}\frac{\partial^2 E_{\pm}}{\partial t^2}\Big). 	
\label{eq:sigma0v2} 
\end{align}

Lastly, it is possible to calculate $ \sigma_{2\pm}$ from equation (\ref{eq:sigma2-v1}) in a similar way:
\begin{align*}
\sigma_{2\pm} = &i\frac{\mu T_{2}}{2\hbar} (1 -T_2\frac{\partial}{\partial t}) (E_{\mp} n^{(*)}_{u2})  \\ 
=&i\frac{\mu T_{2}}{2\hbar} \Big(E_{\mp} n^{(*)}_{u2} - T_2\frac{\partial E_{\mp}}{\partial t}n^{(*)}_{u2} - T_2 E_{\mp} \frac{\partial n^{(*)}_{u2}}{\partial t} \Big). 	
\end{align*}
We have kept the term $\frac{\partial n_{u2}}{\partial t}$ since $n_{u2}$ is a complex value and has impact on both the amplitude and the phase of the field envelope. Combining equations (\ref{eq:nu0v2}) and (\ref{eq:nu2v2}) with the previous relation, one can obtain the final expression for $\sigma_{2\pm}$:
\begin{align}
\begin{split}
\sigma_{2\pm} =&- i\frac{\mu^3 T^2_{2}T_{1}T_{g} J }{4\hbar^3 \Big( 1 + \frac{{\abs*{E_+}}^2+{\abs*{E_-}}^2} {E^2_{sat}}  \Big) }  \Big(  E_{\pm}{\abs*{E_{\mp}}}^2 \\ 
&- (T_{2}+T_{g})\frac{\partial E_{\pm}}{\partial t}{\abs*{E_{\mp}}}^2   
 - (T_{2}+T_{g}) E_{\pm} E_{\mp}\frac{\partial E^{*}_{\mp}}{\partial t} \\ 
 & - T_2 E_{\pm} E^*_{\mp}\frac{\partial E_{\mp}}{\partial t}  \Big). 	
\label{eq:sigma2v2} 
\end{split}
\end{align}

The final step is to insert relations (\ref{eq:sigma0v2}) and (\ref{eq:sigma2v2}) into the wave equation (\ref{eq:E+-v1}). Upon introducing the unsaturated intensity gain factor $g_0$ in units of $m^{-1}$ as:
\begin{align}
g_{0} = \frac{ \Gamma\mu^2 \omega_0 T_{1}T_{2} J }{\hbar n c \varepsilon_0 L  } , 	
\label{eq:g0} 
\end{align}
we can finally write the master equation for the forward and backward propagating components $E_{\pm}$ of the complex field envelope:
\begin{align}
\begin{split}
\frac{n}{c}\frac{\partial E_{\pm}}{\partial t} &\pm  \frac{\partial E_{\pm}}{\partial z} -ik''\frac{\partial^2E_{\pm} }{\partial t^2} = -\frac{\alpha_w}{2}E_{\pm}  \\ 
&+i\beta ( {\abs*{E_+}}^2+{\abs*{E_-}}^2 )E_{\pm}  \\
& + \frac{g_0 }{ 2\Big(1 + \frac{{\abs*{E_+}}^2+{\abs*{E_-}}^2} {E^2_{sat}} \Big) }  \Big[E_{\pm} - T_{2}\frac{\partial E_{\pm}}{\partial t} + T^2_{2}\frac{\partial^2 E_{\pm}}{\partial t^2}\Big]\\
&- \frac{g_0T_g }{ 2T_1E^2_{sat} \Big(1 + \frac{{\abs*{E_+}}^2+{\abs*{E_-}}^2} {E^2_{sat}} \Big)  }  \Big[E_{\pm}{\abs*{E_{\mp}}}^2  \\ 
&- (T_{2}+T_{g})\frac{\partial E_{\pm}}{\partial t}{\abs*{E_{\mp}}}^2 -(T_{2}+T_{g}) E_{\pm} E_{\mp}\frac{\partial E^{*}_{\mp}}{\partial t}  \\ 
&- T_2 E_{\pm} E^*_{\mp}\frac{\partial E_{\mp}}{\partial t}  \Big].
\label{eq:E+-v2} 
\end{split}
\end{align}

In this way, the system of eight coupled differential equations was reduced down to a system of just two coupled differential equations. They provide a good quantitative tool for a laser which possesses fast gain dynamics. However, even if that is not the case, equation (\ref{eq:E+-v2}) is still useful as a qualitative tool to gain insight into the underlying physics. Concerning the quantitative analysis for the slow gain medium, one could utilize the Taylor expansion approach only to the polarization equations and implement the starting differential equations for the populations and complex field envelopes.

From here it is possible to simplify the equation even more. Let us represent the envelopes of the electric field as:
\begin{align}
E_{\pm} = A_{\pm}e^{i\phi_{\pm}},
\label{eq:aphi}
\end{align}
where $A_{\pm}$ are the strictly real amplitudes and $\phi_{\pm}$ are the phases of the forward and backward propagating components of the complex field. For simplicity, we can also introduce the saturated intensity gain $g(P)$ as:
\begin{align}
g(P) = \frac{  g_{0}  }{ 1 + \frac{{\abs*{E_+}}^2+{\abs*{E_-}}^2} {E^2_{sat}}  } = \frac{  g_{0}  }{ 1 + \frac{P} {P_{sat}}  }  , 	
\label{eq:satg} 
\end{align}
where $P=P_++P_-=\abs*{E_+}^2+{\abs*{E_-}}^2$ is the normalized power and $P_{sat}=E^2_{sat}$.
Inserting equations (\ref{eq:aphi}) and (\ref{eq:satg}) into (\ref{eq:E+-v2})  and grouping real and imaginary terms together yields equations for the amplitudes and phases separately:
\begin{flalign}
\frac{n}{c}\frac{\partial A_{\pm}}{\partial t} &{\pm} \frac{\partial A_{\pm}}{\partial z} =   \frac{ g(P)}{2} \Big[ A_{\pm} - T_2 \frac{\partial A_{\pm}}{\partial t} + T^2_2\frac{\partial^2 A_{\pm}}{\partial t^2} \nonumber \\ 
&- T^2_2 A_{\pm}{\Big( \frac{\partial \phi_{\pm}}{\partial t} \Big) }^2 \Big] - \frac{  g(P)  T_g}{2T_1P_{sat}} \Big[ A_{\pm}A^2_{\mp} \nonumber \\ 
&- (T_{2}+T_{g})\frac{\partial A_{\pm}}{\partial t}A^2_{\mp}  - (2T_{2}+T_{g})\frac{\partial A_{\mp}}{\partial t}A_{\pm}A_{\mp} \Big] \nonumber \\
&+k'' \Big[-2\frac{\partial A_{\pm}}{\partial t}\frac{\partial \phi_{\pm}}{\partial t} -A_{\pm}\frac{\partial^2 \phi_{\pm}}{\partial t^2} \Big]  -\frac{\alpha_w}{2}A_{\pm} , \label{eq:am+-} \\ 
\frac{n}{c}\frac{\partial \phi_{\pm}}{\partial t} &{\pm}\frac{\partial \phi_{\pm}}{\partial z}= \frac{ g(P)}{2} \Big[ - T_2 \frac{\partial \phi_{\pm}}{\partial t}  \nonumber \\
& + 2T^2_2 \frac{1}{A_{\pm}}\frac{\partial A_{\pm}}{\partial t}\frac{\partial \phi_{\pm}}{\partial t}+ T^2_2 \frac{\partial^2 \phi_{\pm}}{\partial t^2} \Big] \nonumber \\
& - \frac{  g(P)  T_g}{2T_1P_{sat}} \Big[ -(T_{2}+T_{g})A^2_{\mp} \frac{\partial \phi_{\pm}}{\partial t} + T_{g}A^2_{\mp} \frac{\partial \phi_{\mp}}{\partial t} \Big]  \nonumber\\
&+k'' \Big[\frac{1}{A_{\pm}}\frac{\partial^2 A_{\pm}}{\partial t^2} - {\Big( \frac{\partial \phi_{\pm}}{\partial t} \Big) }^2 \Big] +\beta(A^2_+ + A^2_-).   \label{eq:phi+-} 
\end{flalign}

We can give some comments now about the influence of different terms. From the equation for the amplitude (\ref{eq:am+-}) it is clear why we needed to keep the second derivative in the Taylor expansion in the derivation of $\sigma_{0\pm}$ in the equation (\ref{eq:sigma0v2}). It can be easily seen that it is the only term that gives feedback from the phases $\phi_{\pm}$ to the amplitudes $A_{\pm}$ (if we neglect the dispersion) through $T^2_2 A_{\pm}{\Big( \frac{\partial \phi_{\pm}}{\partial t} \Big) }^2$, so leaving it out results in chaotic behavior, since the amplitudes become independent of the phases. Furthermore, it is the largest in value, and hence has the largest impact, compared to the other second derivative terms that would emerge from Taylor expansions of $\sigma_{2\pm}$ or $n_{u2}$. 

Moreover, SHB has a crucial role in the intermode laser dynamics, since it is the main effect that induces multimode lasing. It decreases the gain competition between cavity modes and allows a large number of side modes to overcome the lasing threshold. We verified this easily by setting the SHB grating term in the population $n_{u2}=0$, which consequently makes $\sigma_{2\pm}=0$. This alteration results always in single mode laser operation. Concerning the master equation (\ref{eq:E+-v2}), the term that dominantly describes SHB effect is $(T_{2}+T_{n_{u2}}) E_{\pm} E_{\mp}\frac{\partial E^{*}_{\pm}}{\partial t}$ given in the second square bracket. Its exclusion causes a single mode solution to arise. This can be understood better by analyzing the equations for the amplitudes and the phases. In equation (\ref{eq:am+-}), this term is represented by $(2T_{2}+T_{n_{u2}})\frac{\partial A_{\mp}}{\partial t}A_{\pm}A_{\mp}$ which gives the contribution from the amplitude derivative of the opposite propagating envelope. It is even more clear when looking at the phase equation (\ref{eq:phi+-}). There, the SHB term is $T_{n_{u2}}A^2_{\mp} \frac{\partial \phi_{\mp}}{\partial t}$ and it is the only term that results in the crosstalk between the phases $\phi_+$ and $\phi_-$ in the corresponding equations.

Equations (\ref{eq:am+-}) and (\ref{eq:phi+-}) are valid both for amplitude and phase modulated laser. However, in an FM comb, one can assume that the laser dynamics are dominantly described with the phase equation (\ref{eq:phi+-}), since it is the phase that is being modulated. The amplitude is approximately constant and is less important. From this follows that the derivatives of the phases $\phi_{\pm}$ are more significant than the derivatives of the amplitudes $A_{\pm}$. That means that we can write simplified equations in the case of an FM frequency comb, disregarding multiple terms from equations (\ref{eq:am+-}) and (\ref{eq:phi+-}), which we have confirmed via numerical simulations:
\begin{flalign}
\frac{n}{c}\frac{\partial A_{\pm}}{\partial t} {\pm} \frac{\partial A_{\pm}}{\partial z} &=   \frac{ g(P)}{2} \Big[ A_{\pm} - T_2 \frac{\partial A_{\pm}}{\partial t}  - T^2_2 A_{\pm}{\Big( \frac{\partial \phi_{\pm}}{\partial t} \Big) }^2 \Big] \nonumber \\
&  -\frac{\alpha_w}{2}A_{\pm} ,\label{eq:ams+-} \\
\frac{n}{c}\frac{\partial \phi_{\pm}}{\partial t} {\pm}\frac{\partial \phi_{\pm}}{\partial z} &= \frac{ g(P)}{2} \Big[  - T_2 \frac{\partial \phi_{\pm}}{\partial t} - \frac{ T^2_{g}}{T_1}\frac{A^2_{\mp}}{P_{sat}} \frac{\partial \phi_{\mp}}{\partial t} \Big]  \nonumber \\
& - k''\Big( \frac{\partial \phi_{\pm}}{\partial t} \Big) ^2  +\beta(A^2_+ + A^2_-) , \label{eq:phis+-} 
\end{flalign}

Alternatively, we can express the last two equations using the power $P_{\pm}$ instead of the amplitude if we write $E_{\pm} = \sqrt{P_{\pm}}e^{i\phi_{\pm}}$:
\begin{flalign}
\frac{n}{c}\frac{\partial P_{\pm}}{\partial t} + \frac{\partial P_{\pm}}{\partial z} &=   g(P) \Big[ P_{\pm} - \frac{T_2}{2} \frac{\partial P_{\pm}}{\partial t}  - T^2_2 P_{\pm}{\Big( \frac{\partial \phi_{\pm}}{\partial t} \Big) }^2 \Big] \nonumber \\
&  -\alpha_w P_{\pm} ,\label{eq:ams1+-} \\
\frac{n}{c}\frac{\partial \phi_{\pm}}{\partial t} {\pm}\frac{\partial \phi_{\pm}}{\partial z} &= \frac{ g(P)}{2} \Big[  - T_2 \frac{\partial \phi_{\pm}}{\partial t} - \frac{ T^2_{g}}{T_1}\frac{P_{\mp}}{P_{sat}} \frac{\partial \phi_{\mp}}{\partial t} \Big]  \nonumber \\
& - k''\Big( \frac{\partial \phi_{\pm}}{\partial t} \Big) ^2  +\beta(P_+ + P_-) .\label{eq:phis1+-} 
\end{flalign}

\begin{figure*}
	\centering
	\includegraphics[width = 1.\textwidth]{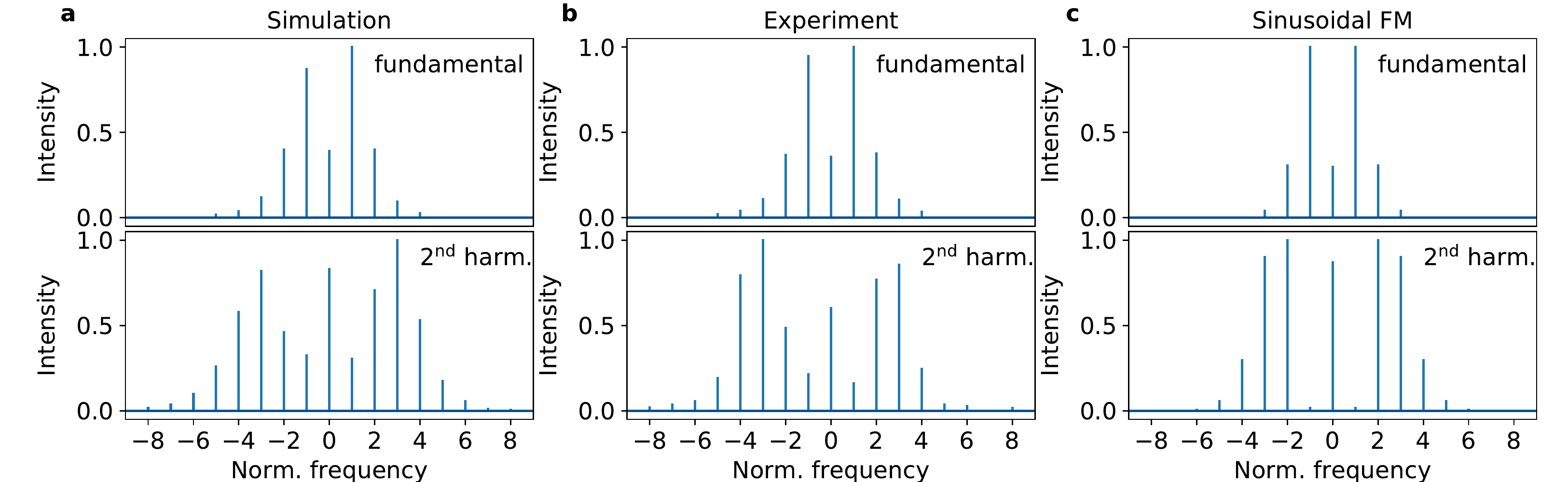}
	\caption{ Modal amplitudes of the fundamental intensity spectra (top row) and the second harmonic spectra (bottom row). (a) Data obtained from our simulation corresponding to a chirped FM. (b) Data recreated from \cite{tiemeijer1989passive}. (c) Bessel amplitudes spectrum, corresponding to a sinusoidal FM. }
	\label{sup_fig1}
\end{figure*}

We will first comment the power equation (\ref{eq:ams1+-}). One sees that the entire second square bracket in the equation (\ref{eq:am+-}) is left out. Terms in that bracket originate from SHB and their influence on the power (amplitude) can be neglected, bearing in mind that we are analyzing the dynamics of an FM comb which is governed by the phase evolution. The term proportional to $( \frac{\partial \phi_{\pm}}{\partial t} ) ^2$ is the only one that gives a feedback from the the phase to the power (amplitude) and is necessary in the case of phase-locking in the presence of the Kerr nonlinearity. Intriguingly, it turns out that it can be left out in the case of phase-locking with GVD. Furthermore, the power can be calculated from the simple equation $\pm \partial P_{\pm}/ \partial z = (g(P)-\alpha_w) P_{\pm}$ and then plugged back in the phase equations. This yields a time independent power with a completely constant output.

We will now turn to the equation (\ref{eq:phis1+-}). The first term in the square brackets leads only to a single mode solution since it dampens the oscillations. The second term in the square brackets is due to SHB and it links together the phases of the opposite propagating fields. This leads to multi-mode operation, proving again that SHB is an absolute necessity. Next, it is seen that both the Kerr nonlinearity and the group velocity dispersion dominantly influence the phase, and not the amplitude of the field. The GVD term proportional to $k''$ shapes the phases into a parabola, which corresponds to linear intermodal phase differences and a chirped frequency. The sign of $k''$ determines whether the parabola is convex or concave, giving rise to a chirp in one direction or another. Concerning the Kerr influence, we can give a following explanation. The second term in the square brackets in the equation (\ref{eq:ams1+-}) is a source term, which drives the power to oscillate in a similar manner as $( \frac{\partial \phi_{\pm}}{\partial t} ) ^2$. Then it is clear that the Kerr term, which proportional to $\beta$ in the equation (\ref{eq:phis1+-}), has a similar impact as the GVD.

\section{Additional results}

\subsection{Chirped vs sinusoidal frequency modulation }

Frequency modulated locked lasers were studied as far back as in the 1960s. First experiments were relying on an active intracavity phase modulator in order to generate FM outputs \cite{Harris1964FM, Kuizenga1970FM}. 
Based on those experiments, the modal amplitudes in the optical spectrum of an FM phase-locked laser were believed to follow a Bessel-function pattern, which is a sign of sinusoidal frequency modulation. Rather unexpected, it was recently discovered that a self-starting FM frequency comb is characterized by a chirped, instead of a sinusoidal frequency modulation \cite{singleton2018evidence,hillbrand2018coherent}.
The reason why this has been overseen for over 50 years most probably lies in the fact that the acquisition of only the laser intensity spectrum is not sufficient to recreate the time trace of the laser output uniquely. One would additionally require the information about spectral phases. However, back then, phase sensitive measurements of the laser output that do not rely on a short pulse emission, have not been developed.

An analysis based only on the comparison of the intensity spectra could provide an indication. However, it can also be misleading. In figure \ref{sup_fig1}b we have shown the intensity spectra measured in \cite{tiemeijer1989passive}. To our knowledge, it was the first observation of a self-starting FM locked laser, obtained from a laser diode, which is a slow gain medium. The authors have concluded that the laser was emitting a sinusoidally modulated frequency output, based on the similarity of the modal amplitudes to the Bessel functions. On figures \ref{sup_fig1}a \& c, we have shown respectively the spectra obtained from the simulation, corresponding to a linearly chirped FM output and Bessel amplitude spectra, corresponding to a sinusoidal FM output. The resemblance of the simulated fundamental spectra to the measured spectrum is much greater than that of the sinusoidal FM. Even in the case of the second harmonic, the measured modal amplitudes correspond better to the ones from our simulation. 
This provides an indication that self-starting FM combs are characterized with a chirped frequency output as a phenomenon general to various types of semiconductor lasers.

\subsection{Impact of spatial hole burning}

When deriving the master equation for FM combs (\ref{eq:E+-v2}), we have isolated the terms in the second square bracket on the right hand side of the equations as contributions from SHB. Keeping this in mind, it becomes easy to analyze the impact of the SHB on the laser dynamics by simply leaving this term out of the equation. 

\begin{figure*}
	\centering
	\includegraphics[width = 1.\textwidth]{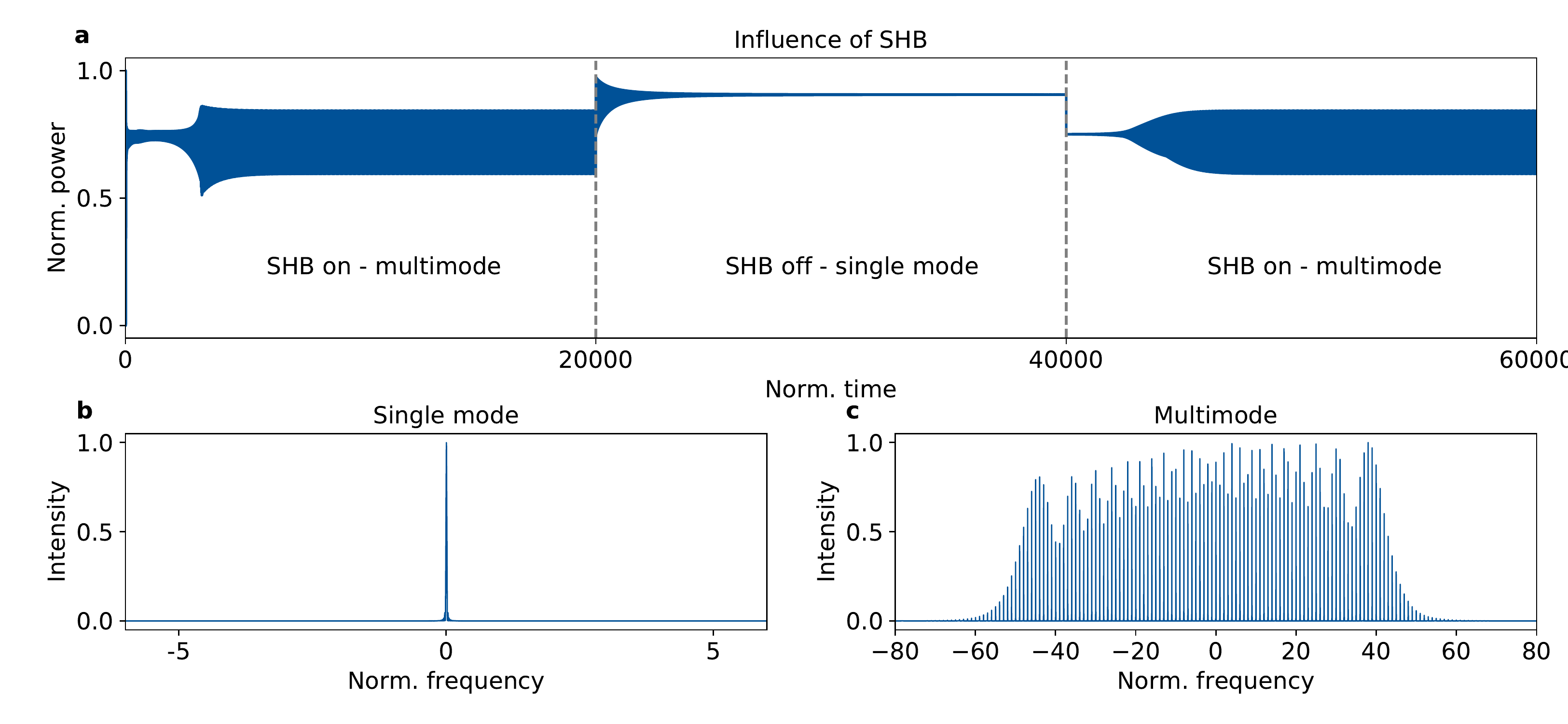}
	\caption{ (a) Normalized output power. Dashed lines limit border regions where the SHB is turned on or off. Time is normalized to the cavity round-trip time. Optical intensity spectra obtained for: (b) single mode operation, when the SHB is turned off and (c) multimode operation, with present SHB }
	\label{sup_fig2}
\end{figure*}

In figure \ref{sup_fig2} we have shown the simulation results of 60000 round-trips of laser operation. The SHB term was turned off after 20000 and turned on back again after 40000 round-trips. The time trace signal of the normalized output power is depicted in figure \ref{sup_fig2}a. The laser becomes phase-locked via GVD after around 10000 round-trips with the intensity spectrum consisted of many modes shown in figure \ref{sup_fig2}c. After switching the SHB off, the laser stabilizes to single mode operation with spectrum shown in figure \ref{sup_fig2}b. Furthermore, the power increases due to the cancellation of the SHB induced losses. After switching the SHB on again, the laser returns to its original state. As a conclusion, carrier grating induced by SHB is necessary (but not sufficient) for FM phase-locking with a linear chirp.

\section{Parameter values}

\hspace{-0.25cm}%
	\begin{tabular}{ |p{1cm}||p{5cm}||p{2.cm}|  }
		\hline
		Sym.& Description & Value\\
		\hline
		$T_{ul}$ & Upper-lower transition time   & $0.5 \: \mathrm{ps} $ \\
		$T_{ug}$ & Upper-ground transition time     & $3 \: \mathrm{ps} $ \\
		$T_{lg}$ & Lower-ground transition time     & $0.08 \: \mathrm{ps} $ \\
		$T_{2}$ & Dephasing time  & $50 \: \mathrm{fs} $ \\
		$n$ & Refractive index  & $3.3 $ \\
		$D$ & Diffusion coefficient  & $46\: \mathrm{cm}^2/s $ \\
		$\alpha_{w}$ & Waveguide power losses   & $4 \: \mathrm{cm}^{-1} $ \\
		$\mu$ & Dipole matrix element  & $2.3 \: \mathrm{nm}\times\mathrm{e}  $ \\
		$n_{tot}$ & Sheet density  & $6\!\times10^{10} \: \mathrm{cm}^{\!-2}  $ \\
		$R_l, R_r$ & Terminal facets reflectivity   & $0.3 $ \\
		$\Gamma$ & Confinement factor   & $1 $ \\
		$L$ & Period length   & $580 \: \angstrom $ \\
		$L_c$ & Cavity length   & $4 \: \mathrm{mm} $ \\
		$\lambda_0$ & Central wavelength   & $8 \: \mathrm{\mu m} $ \\
		\hline
	\end{tabular}

\end{document}